\documentclass[12pt,a4paper]{article}

\def\bb{\begin{equation}}
\def\ee{\end{equation}}
\def\pt{\partial}

\def\const{\hbox{const}}

\def\ve{\varepsilon}

\title{\bf{Soliton generation by local resonance interaction}
\footnote{This work was supported by RFBR grant (00-15-96038)}}
\author{Glebov S.G.,\footnote{Ufa State Petroleum Technical
University; gloomy69@ufanet.ru} \and Kiselev O.M.,
\footnote{Institute of Mathematics  Ufa Centre RAS; ok@ufanet.ru}
\and Lazarev V.A.\footnote{Ufa State Petroleum Technical
University; lazva@mail.ru} }

\date{\null}
\begin{document}
\maketitle

\section{Introduction}

A wide class of the mathematical physics problems   is connected
with forced nonlinear Schrodinger equation (NLSE). Usually
authors  consider either the small amplitude solutions \cite{akh,
klen} or  the asymptotics for perturbed soliton solutions
\cite{Kaup,Karpman-Maslov,Kalyakin}. The problems on soliton
generation usually are not discussed or solved numerically
\cite{barash,frnsh}.
\par
In this paper we show a mechanism of the soliton generation in
nonlinear Schrodinger equation due to a small fast oscillating
driving  force:
\bb i \pt_\theta \Psi + \pt_{\xi\xi} \Psi + |\Psi|^2 \Psi = \ve^2
f e^{iS/\ve^2},\qquad 0< \ve \ll 1. \label{sh}
\ee
We investigate  formal asymptotic solutions with a small
amplitude of the order of ${\cal O}(\ve^2)$. The leading-order
term  is the solution of linear nonhomogeneous equation. We
consider the one phase asymptotic solution with the phase
function $S/\ve^2$. Other terms of the order of ${\cal O}(\ve^2)$
do not play any role in investigated phenomenon.
\par
The specific of the problem consists in existence of
 a local resonance. The resonance curve  is defined by
equation:
\bb
\pt_\theta S + (\pt_\xi S)^2 = 0, \label{aikonal}
\ee
where $S$ is the phase function of the driving  force. The
amplitude of forced oscillations increases from ${\cal O}(\ve^2)$
up to ${\cal O}(\ve)$ in the neighborhood of this curve.
\par
The increase of the amplitude is similar to a phenomenon of a
local resonance  in the linear nonhomogeneous ordinary
differential equation of the second order with the constant
coefficients \cite{kev1}. The amplitude increases linearly. The
domain of this increase usually is called a resonance layer. The
additional term of the order of the layer width will appear in the
asymptotic solution after passage of the resonance layer. The
order ${\cal O}(\ve^2)$ of residual part of the solution does not
change.
\par
After resonance the leading-order term of the asymptotic solution
has the order of ${\cal O}(\ve)$. The spatial variable $\ve\xi$
is defined by amplitude of the driving force. The relation
between the order of the solution amplitude  and typical scale of
the independent variables is such that the leading-order term of
the asymptotics is defined from NLSE. So specific relations
usually appear due to specific asymptotic representation of the
solution as was shown in well known works on asymptotic transfer
to NLSE \cite{klen}, \cite{zakh}-\cite{kel}. In this work we show
that such kind relation can be obtained due to resonance with
driving force.
\par
Thus after resonance layer the leading-order term of the
asymptotics has the order ${\cal O}(\ve)$ and is defined from the
Cauchy problem. The initial data are equal to the value of the
amplitude $f$ of driving  force on the resonance curve. The
further evolution of the leading term of the asymptotics does not
depend on driving  force.
\par
The effect of the solution increase after local resonance is well
known in the ordinary differential equations \cite{kev1} and
partial differential equations \cite{lk2,glebskie}. Also it is
well known the effect of appearance of NLSE in asymptotic
constructions for the problems with strong dispersion, cubic
nonlinearity and  the appropriate relation between scales. This
problem contains both of these effects. The passage through the
local resonance leads to the increase of the amplitude. The
modulation of the amplitude of the driving  force and scales of
the variables lead to NLSE behind the resonance curve. The similar
results can be obtained for other  problems on the resonance
passage by driving force in systems  with the cubic nonlinearity
and strong dispersion.
\par
The considered problem is the model problem for the wide class
problems on the  local resonance passage by driving force. In
order to show essence of this phenomenon we assume that the
amplitude $f$ of perturbation is a smooth function with respect to
variable $\ve \xi$ and phase function is $S=(\ve^2 \theta)^2/2$.
\par
The result of this work is the following proposition

\noindent {\bf Proposition.} {\it One phase asymptotic solution
of the order of ${\cal O}(\ve^2)$ in the domain $\theta\ll
-\ve^{-1}$ will have the order of ${\cal O}(\ve)$ in the domain $
\ve^{-1} \ll \theta \le K\ve^{-2}, \ K=const >0$. The
leading-order term of the asymptotics
$$
\Psi=\ve \stackrel{0}{u} +{\cal O}(\ve^2)
$$
is determined from the Cauchy problem for NLSE
\begin{eqnarray*}
i\stackrel{0}{u}_t+\stackrel{0}{u}_{xx}+
|\stackrel{0}{u}|^2\stackrel{0}{u}=0,\\
\stackrel{0}{u}|_{t=0}=(1-i)\sqrt{\pi} f(x).
\end{eqnarray*} }
\par
In particular, if the amplitude of the perturbation force has the
form
$$
f(x)={1\over(1-i)\sqrt{\pi}}{2\eta\exp(-i\mu (x+ y_0))\over
\cosh(2\eta (x -  x_0))},
$$
then the leading-order term $\stackrel{0}{u}$ of postresonance
expansion is a soliton of NLSE, where $x_0, y_0, \mu, \eta$ are
soliton parameters.
\par

\section{Forced oscillations}

In this section we construct the formal asymptotic solution for
equation (\ref{sh}) in preresonance domain.
\par
{\bf Lemma 1.}{\it \ In the domain  $\theta \ll -\ve^{-1}$ the
formal asymptotic solution for equation (\ref{sh}) with respect
to base ${\cal O}(\ve^{8})$ has a form
\begin{eqnarray}
\Psi(\xi,\theta,\ve) =
\left[\ve^2\stackrel{1}{A}(\ve\xi,\ve^2\theta)+
\ve^{4}\stackrel{2}{A}(\ve\xi,\ve^2\theta)+\right. \label{asex} \\
\left.  \ve^{6}\stackrel{3}{A}(\ve\xi,\ve^2\theta)\right]
\exp(iS/\ve^2), \quad  \ve \to 0, \nonumber
\end{eqnarray}
the coefficients $\stackrel{m}{A}$ of the asymptotics are
determined from algebraic equations (\ref{algex}).}
\par
Equation (\ref{sh}) is written in terms of fast variables $\xi$ è
$\theta$ but further for convenience  we also use slow variables
$x=\ve \xi$ and $t=\ve^2 \theta$.
\par
The substitution (\ref{asex}) into equation (\ref{sh}) gives  a
recurrent sequence of equations for coefficients of asymptotics:
\begin{eqnarray}
-S'\stackrel{1}{A}=f, \nonumber \\  -S'\stackrel{2}{A}=-i
\stackrel{1}{A}_t - \pt_{xx}\stackrel{1}{A}, \label{algex} \\
-S'\stackrel{3}{A}= -i\stackrel{2}{A}_t - \pt_{xx}\stackrel{2}{A}
- |\stackrel{1}{A}|^2\stackrel{1}{A}. \nonumber
\end{eqnarray}
The solutions of this systems have singularities at $t=0$. The
constructed asymptotic solution can be represented  as $t\to0$ in
the form:
$$
\psi(\ve\xi,t,\ve)=\left[\ve^{2}\bigg({f\over t}\bigg)+
\ve^{4}\bigg({if\over t^3}-{f_{xx}\over t^2}\bigg)+\right.
$$
$$
\left.\ve^{6}\bigg({3if_{xx}\over t^4} -{f_{xxxx}\over t^3}
-{|f|^2 f\over t^4} + {3f\over t^5}\bigg)\right]\exp(iS/\ve^2).
$$
The structure of singularities as $t \to -0$  of coefficients in
asymptotic representation (\ref{asex}) allows us to determine the
domain of validity for external asymptotics:
$$
\theta \ll -\ve^{-1}.
$$
Lemma 1 is proved.
\par

\section{Internal solution}

In this section we construct the formal asymptotic solution for
equation (\ref{sh}) into resonance layer. The asymptotics of these
type are usually called by internal asymptotics \cite{ilin}.
\par
{\bf Lemma 2.}{\it \ In the domain  $ -\ve^{-2} \ll \theta \ll
\ve^{-2}$ the formal asymptotic solution for equation (\ref{sh})
with respect to base ${\cal O}(\ve^{4})$ has a form
\bb
\Psi(\xi,\tau,\ve)=\ve\stackrel{0}{w}(\ve\xi,\tau)+
\ve^{2}\stackrel{1}{w}(\ve\xi,\tau)+
\ve^3\stackrel{2}{w}(\ve\xi,\tau) \qquad
 \ve \to 0, \label{internal-anzats}
\ee
where $\tau=\ve \theta$. The coefficients $\stackrel{m}{w}$ of the
asymptotics are determined from ordinary differential equations
(\ref{inw0}), (\ref{inw1}), (\ref{inw2}).}
\par
The structure of singularities for coefficients in (\ref{asex}) as
$t \to -0$ allows us to determine internal scaled variable $\tau
= t /\ve=\ve \theta$.
\par
The internal asymptotic solution is constructed in form
(\ref{internal-anzats}). The behaviour of coefficients  can be
obtained by matching:
\bb
\stackrel{0}{w}(\ve\xi,\tau)\sim \left[-{f\over \tau}+{if\over
\tau^3} +{3f\over \tau^5}\right]\exp\{i\tau^2/2\},\quad \tau\to
-\infty.\label{w0-as-left}
\ee
\bb
\stackrel{1}{w}(\ve\xi,\tau)\sim \left[ -{f_{xx}\over \tau^2}+
{3if_{xx}\over \tau^4}-{|f|^2f\over
\tau^4}\right]\exp\{i\tau^2/2\} ,\quad \tau\to
-\infty.\label{w1-as-left}
\ee
\bb
\stackrel{2}{w}(\ve\xi,\tau)\sim \left[-{f_{xxxx}\over
\tau^3}\right]\exp\{i\tau^2/2\},\quad \tau\to
-\infty.\label{w2-as-left}
\ee
\par
The leading-order term of (\ref{internal-anzats}) is determined
from equation:
\begin{eqnarray}
i\stackrel{0}{w}_\tau=f\exp(i\tau^2/2). \label{inw0}
\end{eqnarray}
The solution of this equation can be written out in terms of
Fresnel integral:
\bb
\stackrel{0}{w}=-if(\ve\xi)\int_{-\infty}^\tau\exp(i\theta^2/2)d\theta.
\label{leading-order-of internal-exp}
\ee
The first order correction $\stackrel{1}{w}$ is determined by
equation:
\bb
i\stackrel{1}{w}_\tau+\stackrel{0}{w}_{xx}+
|\stackrel{0}{w}|^2\stackrel{0}{w}=0, \label{inw1}
\ee
the second order correction is determined by equation:
\bb
i\stackrel{2}{w}_\tau+\stackrel{1}{w}_{xx}+
2|\stackrel{0}{w}|^2\stackrel{1}{w}+
\stackrel{0}{w}{}^2\stackrel{1}{w}{}^*=0. \label{inw2}
\ee
\par
In order to determine the behaviour of the asymptotic solution
after resonance passage we need to know the asymptotics  of the
coefficients of (\ref{internal-anzats}) as $\tau \to +\infty$.
Direct calculations lead us to formulas
$$
\stackrel{0}{w}(\ve\xi,\tau)=-i f(\ve\xi)\bigg[ic_1 +
{\exp(i\tau^2/2)\over i\tau}+{\cal O}(\tau^{-3})\bigg],
$$
where $c_1 = (1-i)\sqrt{\pi}$;
$$
\stackrel{1}{w}(\ve\xi,\tau)=\tau\stackrel{1}{w}{}_{1}(\ve\xi) +
\stackrel{1}{w}{}_0(\ve\xi)+ g_1(\ve\xi){\exp(i\tau^2/2)\over
i\tau^2} +{\cal O}(\tau^{-4}),
$$
where
$$
\stackrel{1}{w}{}_1=c_1f_{xx}+|c_1 f|^2 c_1 f,
$$
$$
\stackrel{1}{w}{}_0(\ve\xi)=\lim_{\tau\to\infty}\bigg(
\int_{-\infty}^\tau \big[\stackrel{0}{w}{}_{xx}(\ve\xi,\theta)+
|\stackrel{0}{w}(\ve\xi,\theta)|^2\stackrel{0}{w}(\ve\xi,\theta)\big]d\theta
-(c_1f_{xx}+|c_1 f|^2 c_1 f)\bigg),
$$
$g_1(\ve\xi)=k_1 f_{xx}+k_2|f|^2 f$, $k_1$ and $k_2$ are
constants;
$$
\stackrel{2}{w}(\ve\xi,\tau)=\tau^2\stackrel{2}{w}{}_{2}(\ve\xi)
+ \tau\stackrel{2}{w}{}_1(\ve\xi) +\stackrel{2}{w}{}_0(\ve\xi)+
g_2(\ve\xi){\exp(i\tau^2/2)\over i\tau}+{\cal O}(\tau^{-2}),
$$
where
$$
\stackrel{2}{w}{}_2(\ve\xi)=i\left((\stackrel{1}{w}_1)_{xx}
+2|\stackrel{0}{w}_0|^2\stackrel{1}{w}_1 +
\stackrel{0}{w}{}_0^2\stackrel{1}{w}{}_1^*\right),
$$
\begin{eqnarray*}
\stackrel{2}{w}{}_1(\ve\xi)= i\left((\stackrel{1}{w}{}_0)_{xx}+
2|\stackrel{0}{w}_0|^2\stackrel{1}{w}_0+ 2\stackrel{0}{w}{}_0
\stackrel{0}{w}{}_{-1}^*\stackrel{1}{w}_1- \right. \\
\left. -2\stackrel{0}{w}_{-1}
\stackrel{0}{w}{}_0^*\stackrel{1}{w}_1 +2\stackrel{0}{w}{}_0
\stackrel{0}{w}{}_{-1}\stackrel{1}{w}{}_1^* +
\stackrel{0}{w}{}_0^2\stackrel{1}{w}{}_0^*\right),
\end{eqnarray*}
\begin{eqnarray*}
\stackrel{2}{w}{}_0=\lim_{\tau\to\infty}\bigg(
i\int_{-\infty}^\tau \big[\stackrel{1}{w}{}_{xx}(\ve\xi,\theta)+
2|\stackrel{0}{w}(\ve\xi,\theta)|^2\stackrel{1}{w}(\ve\xi,\theta)
+\stackrel{0}{w}{}^2\stackrel{1}{w}^* \big]d\theta\\
-\tau^2\stackrel{2}{w}_2- \tau\stackrel{2}{w}_1\bigg)
\end{eqnarray*}
and the function $g_2(\ve\xi)$ is smooth and expressed by
$f(\ve\xi)$.
\par
The asymptotic behaviour of the coefficients in
(\ref{internal-anzats}) gives us the domain of validity for this
solution:
$$
|\tau| \ll \ve^{-1} \quad \hbox{or} \quad |\theta| \ll \ve^{-2}
$$
Lemma 2 is proved.
\par

\section{Postresonance solution}

In this section we construct the formal asymptotic solution for
equation (\ref{sh}) in postresonance domain.
\par
{\bf Lemma 3.}{\it \ In the domain  $\ve^{-1} \ll \theta \le K
\ve^{-2}$ the formal asymptotic solution for equation (\ref{sh})
with respect to base ${\cal O}(\ve^{4})$ has a form
\begin{eqnarray}
\Psi(\xi,\theta,\ve)=\ve \stackrel{0}{u}(\ve\xi,\ve^2\theta)+
\ve^{2}\big(\stackrel{1}{u}(\ve\xi,\ve^2\theta)+
\stackrel{1}{B}(\ve\xi,\ve^2\theta)\exp(iS/\ve^2)\big)+
\nonumber\\ \ve^3\big(\stackrel{2}{u}(\ve\xi,\ve^2\theta)+
\stackrel{2}{B}_1(\ve\xi,\ve^2\theta)\exp(iS/\ve^2)+
\stackrel{2}{B}_{-1}(\ve\xi,\ve^2\theta)\exp(-iS/\ve^2)\big)+
\label{external-anzats-2}\\
+\ve^{4}\big(\stackrel{3}{B}_1(\ve\xi,\ve^2\theta)\exp(iS/\ve^2)+
\stackrel{3}{B}_2(\ve\xi,\ve^2\theta)\exp(2iS/\ve^2)+\nonumber\\+
\stackrel{3}{B}_{-1}(\ve\xi,\ve^2\theta)\exp(-iS/\ve^2)\big),\nonumber
\end{eqnarray}
the coefficients $\stackrel{m}{B}_k$ of the asymptotics are
determined from algebraic equations (\ref{alg2}) and  the
coefficients $\stackrel{m}{u}$ are solutions of the Cauchy
problems  for either NLSE or  linearized  NLSE
 (\ref{problem-for-u0})-(\ref{problem-for-u2}).}
\par
The substitution of (\ref{external-anzats-2}) in equation
(\ref{sh}) gives:
\begin{eqnarray*}
\ve\bigg(-S'\stackrel{1}{B}-f\bigg)\exp(iS/\ve^2)+\ve^2
\bigg(i\stackrel{0}{u}_t+\stackrel{0}{u}_{xx}+
|\stackrel{0}{u}|^2\stackrel{0}{u}\bigg)
\\
+\ve^{3}\bigg(i\stackrel{1}{u}_t+\stackrel{1}{u}_{xx}
+2|\stackrel{0}{u}|^2\stackrel{1}{u}+
\stackrel{0}{u}{}^2\stackrel{1}{u}{}^*+\\
+i\stackrel{1}{B}_t\exp(iS/\ve^2)+
\stackrel{1}{B}_{xx}\exp(iS/\ve^2)+\\
+2|\stackrel{0}{u}|^2\stackrel{1}{B}\exp(iS/\ve^2)+
\stackrel{0}{u}{}^2\stackrel{1}{B}{}^*\exp(-iS/\ve^2) -\\
-S'\stackrel{2}{B}_{1}\exp(iS/\ve^2)+
S'\stackrel{2}{B}_{-1}\exp(-iS/\ve^2)\bigg)\\
+\ve^4\bigg( i\stackrel{2}{u}_t+\stackrel{2}{u}_{xx}
+2|\stackrel{0}{u}|^2\stackrel{2}{u}+
\stackrel{0}{u}{}^2\stackrel{2}{u}{}^*
+2|\stackrel{1}{u}|^2\stackrel{0}{u} +
\stackrel{1}{u}{}^2\stackrel{0}{u}{}^* +
2\stackrel{0}{u}|\stackrel{1}{B}|^2+\\
+\stackrel{0}{u}{}^*\stackrel{1}{B}{}^2\exp(2iS/\ve^2)+
2\stackrel{0}{u}{}^*\stackrel{1}{u}\stackrel{1}{B}\exp(iS/\ve^2)+\\
+2\stackrel{0}{u}\stackrel{1}{u}{}^*\stackrel{1}{B}\exp(iS/\ve^2)+
2\stackrel{0}{u}\stackrel{1}{u}\stackrel{1}{B}{}^*\exp(-iS/\ve^2)+\\
+2|\stackrel{0}{u}|^2\stackrel{2}{B}_{1}\exp(iS/\ve^2)+
2|\stackrel{0}{u}|^2\stackrel{2}{B}_{-1}\exp(-iS/\ve^2)+\\
+\stackrel{0}{u}{}^2\stackrel{2}{B}{}_{1}^*\exp(-iS/\ve^2)+
\stackrel{0}{u}{}^2\stackrel{2}{B}{}_{-1}^*\exp(iS/\ve^2)+\\
+i\stackrel{2}{B}_{1}{}_t\exp(iS/\ve^2)+
\stackrel{2}{B}_{1}{}_{xx}\exp(iS/\ve^2)-\\
-i\stackrel{2}{B}_{-1}{}_t\exp(-iS/\ve^2)+
\stackrel{2}{B}_{-1}{}_{xx}\exp(-iS/\ve^2)+\\
- S'\stackrel{3}{B}_{1}\exp(iS/\ve^2)+
S'\stackrel{3}{B}_{-1}\exp(-iS/\ve^2) +
2S'\stackrel{3}{B}_{2}\exp(2iS/\ve^2)\bigg) = {\cal O}(\ve^5)
\end{eqnarray*}
\par
Functions  $\stackrel{1}{B}$, $\stackrel{2}{B}_{\pm}$ are
determined from algebraic equation:
\begin{eqnarray}
-S'\stackrel{1}{B}=f,\nonumber \\
S'\stackrel{2}{B}_1=i\stackrel{1}{B}_{t}+\stackrel{1}{B}_{xx}+
|\stackrel{0}{u}|^2\stackrel{1}{B}, \nonumber\\
S'\stackrel{2}{B}_{-1}=-\stackrel{0}{u}{}^2 \stackrel{1}{B}{}^*\nonumber\\
S'\stackrel{3}{B}_1=i\stackrel{2}{B}_1{}_{t}+\stackrel{2}{B}_1{}_{xx}+
|\stackrel{0}{u}|^2\stackrel{1}{B}+
2\stackrel{0}{u}{}^*\stackrel{1}{u}\stackrel{1}{B}+
\stackrel{0}{u}\stackrel{1}{u}{}^*\stackrel{1}{B}+
\stackrel{0}{u}{}^2\stackrel{2}{B}{}_{-1}^*,\label{alg2} \\
S'\stackrel{3}{B}_{-1}=i\stackrel{2}{B}_{-1}{}_{t}-\stackrel{2}{B}_{-1}{}_{xx}
-\stackrel{0}{u}\stackrel{1}{u}\stackrel{1}{B}{}^*-
\stackrel{0}{u}{}^2\stackrel{2}{B}{}_{1}^*,\nonumber\\
2S'\stackrel{3}{B}_2=\stackrel{0}{u}{}^*\stackrel{1}{B}{}^2,
\nonumber
\end{eqnarray}
The matching with internal solutions give us the recurrent system
of the Cauchy problems for coefficients $\stackrel{0}{u},
\stackrel{1}{u}, \stackrel{2}{u}$:
\begin{eqnarray}
i\stackrel{0}{u}_t+\stackrel{0}{u}_{xx}+
|\stackrel{0}{u}|^2\stackrel{0}{u}=0,\nonumber\\
\stackrel{0}{u}|_{t=0}=c_1 f(x);\label{problem-for-u0}\\
i\stackrel{1}{u}_t+\stackrel{1}{u}_{xx}+
2|\stackrel{0}{u}|^2\stackrel{1}{u}+
\stackrel{0}{u}{}^2\stackrel{1}{u}{}^*=0,\nonumber\\
\stackrel{1}{u}|_{t=0}=\stackrel{1}{w}_0(x);\label{problem-for-u1}\\
i\stackrel{2}{u}_t+\stackrel{2}{u}_{xx}+
2|\stackrel{0}{u}|^2\stackrel{2}{u}+
\stackrel{0}{u}{}^2\stackrel{2}{u}{}^*=-|\stackrel{1}{B}|^2\stackrel{0}{u}
-|\stackrel{1}{u}|^2\stackrel{0}{u}-\stackrel{1}{u}{}^2\stackrel{0}{u}{}^*,\nonumber\\
\stackrel{2}{u}|_{t=0}=\stackrel{2}{w}_0(x).\label{problem-for-u2}
\end{eqnarray}
\par
Coefficients $\stackrel{j}{B}_{k}$ have singularities at $t=0$.
The domain of validity for solution (\ref{external-anzats-2}) is
determined by inequalities:
$$
t\gg\ve \quad \mbox{or} \quad \theta \gg \ve^{-1}.
$$
Lemma 3 is proved.
\par
In order to determine the behaviour of solution
(\ref{external-anzats-2}) for large $t>0$ it is necessary to know
the structure of the leading-order term $\stackrel{0}{u}$. It is
easy to see that the behaviour of $\stackrel{0}{u}$ is determined
by amplitude of the perturbation force in original equation
(\ref{sh}). In particular, if the function  $f(\ve \xi)$ has the
form:
$$
f(\ve \xi)={1\over(1-i)\sqrt{\pi}}{2\eta\exp(-i\mu (\ve \xi+
y_0))\over \cosh(2\eta (\ve \xi -  \xi_0))},\quad
\hbox{where}\,\,\, \xi_0, y_0, \mu, \eta=\const,
$$
then to determine the leading-order term of asymptotics
(\ref{external-anzats-2}) it is necessary to solve the Cauchy
problem for NLSE with initial data of the soliton type.
\par
We are grateful to L.A. Kalyakin and B.I. Suleimanov for
stimulating discussions and for help in improving of the
mathematical presentation of the results.
\par

\end{document}